\documentclass{article}
\usepackage{fullpage} 
\usepackage[utf8]{inputenc}
\usepackage{newunicodechar}
\usepackage{graphicx}

\usepackage[T1]{fontenc}
\usepackage{textcomp}
\usepackage{mathptmx}

\let\oldnewcommand=\newcommand
\let\newcommand=\providecommand
\usepackage{amsmath}
\let\newcommand=\oldnewcommand

\usepackage{setspace}                    
\usepackage[usenames]{color}
\usepackage{boxedminipage} 
\usepackage{fancyvrb}  
\usepackage{relsize} 
\usepackage{xcolor}
\usepackage{listings}
\usepackage[printwatermark]{xwatermark}

\usepackage{wrapfig} 

\usepackage{enumitem} 

\usepackage{hyperref} 

\begin{document}


\lstset{language=sh,basicstyle=\ttfamily}

\title{On the security of ballot marking devices}
\author{Dan S. Wallach\thanks{
Professor, Department of Computer Science; Rice Scholar, Baker
    Institute for Public Policy; Rice University; dwallach@rice.edu}}
\date{December 12, 2019\thanks{Original version released August 5,
    2019. \url{https://arxiv.org/abs/1908.01897} The paper in its
  final form will be published by {\em The Ohio State Technology Law
    Journal}.}}
\maketitle

\begin{abstract}
A recent debate among election experts has considered whether electronic
ballot marking devices (BMDs) have adequate security against the risks of
malware. A malicious BMD might produce a printed ballot that disagrees with
a voter's actual intent, with the hope that voters would be unlikely to
detect this subterfuge. This paper considers how an election administrator
can create reasonable auditing procedures to gain confidence that their fleet of
BMDs is operating correctly, allowing voters to benefit from the usability
and accessibility features of BMDs while the overall election still
benefits from the same security and reliability properties we expect
from hand-marked paper ballots.

\end{abstract}

\section{Introduction}
\label{sec:introduction}
Every voting system must protect against a variety of security
threats. It's the essential purpose of any voting system to provide
evidence that its stated outcomes are correct, even in the face of
adversaries who may wish to tamper with it. Every voting system must
also provide usability and accessibility features, because errors in
human voters' operation of the voting system can lead to changes in
the outcome, particularly if the margin of victory is smaller than the
margin of human error.

In the early 2000's, paperless electronic voting systems gained
prominence for their ability to offer important accessibility features
(e.g., optionally large text, button boxes, multiple languages,
headphones), but these systems also created unacceptable security
vulnerabilities. Tampered or even buggy software could corrupt or
destroy all evidence of voters' original intent (see, e.g., Kohno 
et~al.~\cite{kohno04analysis}).

Electronic ballot marking devices (BMDs) would seem to
bridge the gap between the fundamental security properties of paper,
which cannot be overwritten or tampered by any computer and thus
create the potential for elections to be {\em software independent},
and the variety of usability features available with computers, which
cannot be provided in an equivalent manner by
paper-and-pen. BMDs thus have the potential to provide the best of
both worlds.

Recently, Appel, DeMillo, and Stark
(hereafter, ``ADS'')~\cite{ads2019} 
staked out some important security claims, arguing against BMDs.
They argue that voters can easily be fooled and will neither
notice deliberate errors, nor even if they do notice will they
have any meaningful proof of the BMD's misbehavior.

We need to consider exactly how often a voter might notice an error,
what common electoral processes will do next, and how they might be
enhanced. We'll also need to discuss the properties of hand-marked
paper ballots, considered by ADS and many others to be the ``gold
standard'' for election security.

\subsection{Why not just mark ballots by hand?}

A central question, posited by many election integrity activists,
is why we don't just stick with hand-marked paper ballots. This
question is important to address directly.

\begin{itemize}
\item	Not every voter has the ability to do all the tasks necessary to
  read, mark, and cast a paper ballot. Some voters have low vision or zero
  vision. Some voters have limited motor control. Some voters are
  illiterate or dyslexic. Some voters have multiple such issues. BMDs have
  the potential to make voting far more accessible to these
  populations. BMDs can also offer a variety of different languages,
  both in text and voice, offering greater assistance to non-native English
  speakers. Furthermore, {\em Federal and state laws generally make these features
  mandatory}.

\item Ballot marking devices also have the advantage of eliminating
  complete classes of voting errors that can occur with hand marked
  paper ballots. For example, with a BMD it is impossible to
  ``overvote''; the BMD can enforce common rules like ``only one vote
  per contest''. Enforcing such rules is even more important with
  voting methods that allow multiple selections in a contest, such as
  rank-choice or instant-runoff voting. BMDs additionally do not allow
  voters to make stray or ambiguous marks. If a voter needs to change
  their mind after the ballot is printed, they can ``spoil'' it and
  start over again. For contrast, consider the 2008 recount of the
  very close Minnesota Senate race between Coleman and
  Franken. Ambiguous hand-marked ballots were individually considered
  in litigation after the election. 

\item A well-designed BMD can also help every voter to accurately
  convey their intent. For example, a BMD will commonly have a
  confirmation screen at the end of the process that can highlight
  contests that a voter might have accidentally skipped. Features
  like this become even more important as ballots grow longer and more
  complicated. Likewise, a BMD does not face the space constraints of
  a hand-marked paper ballot, allowing each question to appear on a separate
  screen, and thus help prevent voters from accidentally skipping over
  a contest. For contrast, consider the paper ballot in Broward
  County, Florida in 2018, where the contests for U.S. Senate and
  Congressional Representative were placed under the long ballot
  instructions in the left column, leading a potentially significant
  number of voters to miss them entirely\footnote{Appel has written a
    summary of this issue in Florida 2018, alongside other famous
    ballot layout failures. Bad ballot layout can induce high
    undervote rates in any voting technology, but at least BMDs can
    operate without the constraint of compressing a lengthy ballot to fit onto
    a sheet of paper.~\cite{appel-florida2018}}.

\item An important nationwide trend is the consolidation of polling
  places, both for early voting and on election days. Such ``vote
  centers'' allow any voter to cast any one of potentially thousands
  of distinct ballot styles. To run a vote center with hand-marked
  paper ballots, this requires having laser printers for ``ballot on
  demand'' printing. Unfortunately, laser printers have their own
  issues. Most notably, they require a significant power draw (i.e.,
  several kilowatts) to warm up the toner drum, which can cause
  problems in buildings with older wiring. For the same reasons, laser
  printers cannot operate on consumer-grade UPS batteries. If the
  power goes out, the election is dead in the water. Conversely, BMDs
  generally use thermal printers, which are low power and have no
  consumables like ink or toner cartridges. Commercial BMDs have (or
  should have) enough battery to run for hours without
  power. BMD-based elections will be more robust in the face of power
  failures.

\end{itemize}

\noindent
Consequently, a fundamental challenge we face in any BMD
implementation is trying to combine the security properties of
hand-marked paper ballots with the usability and operational benefits
of a BMD.
In Section~\ref{sec:bmd}, we try to define, exactly, what is a BMD.
In Section~\ref{sec:threat}, we consider how we might model the threat
and what we might hope to accomplish with an audit.
In Section~\ref{sec:live-audit}, we describe a ``live auditing''
process and analyze how likely it might be to detect misbehaving BMDs.
In Section~\ref{sec:non-uniform}, we consider attackers' ability
to hide malicious behavior and the ability for an election official
to detect it, such as by observing anomalously high rates of spoiled ballots.
In Section~\ref{sec:tactical}, we consider how an election official
might be able to tactically improve their chances of catching malware.
In Section~\ref{sec:additional}, we consider a variety of other
arguments that have been made, including the question of why
BMDs might or might not be preferable to hand-marked paper ballots.
We conclude in Section~\ref{sec:conclusions}.

\section{What, exactly, is a BMD?}
\label{sec:bmd}
Fundamentally, a BMD is a device that knows about all the different {\em ballot
styles} that a voter might see. By inserting an unfilled ballot, or perhaps
a blank sheet of paper with only a barcode indicating the specific ballot
style, the BMD can then present a touch-screen interface to the voter to
select their choices. BMDs typically include a variety of accessibility
features (button boxes, headphones with audio output, font size settings,
and other features as a supplement to the touch-screen), allowing a larger
number of voters to operate these devices without assistance. When the
voter is finished, a BMD does what its name says: it prints a marked
ballot. 

After that, BMDs come in two varieties: {\em stateful} and
{\em stateless}. The former retains an electronic copy of every
ballot, while the latter promptly forgets what it saw and starts over
again. Stateful BMDs might allow for faster tallies, and provide
redundancy against catastrophic failures (e.g., lost ballot
boxes). Stateless BMDs might be simpler to construct, and provide
stronger guarantees against the impact of malware within the machine.

What happens with the paper ballots after they're printed varies from
vendor to vendor. Typically, the voter will carry it by hand to a
ballot box, which then has a scanner on top. For hand-marked ballots,
these scanners can flag common error modes, including when a voter has
indicated more than one vote in a contest which allows at most one vote
(i.e., ``overvoting'', which can never happen with a BMD, but is
possible with hand-marked paper). Some vendors offer ``privacy
sleeves'' to allow poll workers to do this operation on behalf of
voters who do not have the necessary manual dexterity, while still
preserving the voter's privacy. One system, Los Angeles County's
VSAP\footnote{\url{https://vsap.lavote.net/}}, has the printer and
ballot box integrated together, so that the entire process can be
completed independently and privately.

Integrating the printing and casting would seem to have desirable
usability properties, particularly for voters with limited manual
dexterity. However, such BMDs creates additional security concerns, where a
BMD might print something contrary to the voter's stated desires and
automatically cast it without any opportunity for the voter to
intervene~\cite{appel-expressvote-xl}.

The rest of this paper will focus primarily on stateless BMDs, wherein
{\em the paper ballot is the only way to know the voter's intent}, and
where {\em vote casting is a manual process, where the voter moves the
  ballot to a physically distinct ballot box.}  This simplifies the
discussion, and makes it clear exactly what ``the ballot'' actually
is. In particular, this makes it clear what happens during a recount,
where ``recounting the ballots'' means looking at paper ballots, not
electronic records. This also simplifies our discussion of what it
means to ``cast'' a ballot.

\section{Threat and audit models}
\label{sec:threat}
Every BMD is just a computer. Like any computer, it might have
bugs in its software that don't turn up in testing and might then impact
the voter's experience. Also, like any computer, its software could include
{\em malware}, not intended by the manufacturer or election official to be
present, but perhaps surreptitiously inserted when nobody was
looking. Plenty of opportunities for this exist in modern elections, where
voting machines may be delivered days or weeks in advance of an
election. (This is colloquially referred to as the ``sleepover voting
machine problem''. A variety of physical security protocols have been
deployed to mitigate these threats, but this is beyond the scope of
this paper.)

In the mid-2000's, researchers were concerned with how this sort of attack
might play out with paperless electronic voting systems (which typically
went by the unwieldy acronym ``direct recording electronic''---DRE), since
a voting machine might {\em appear} to be operating correctly, displaying
exactly what the voter intended, but secretly record the ballot internally
in a very different fashion. A related issue is that a paperless electronic
system can also retain the ballots {\em in the order cast}, or randomize
them in a reversible fashion, allowing ballot secrecy to be compromised by
anybody who observers the order in which voters arrive at the polls.

The mitigations that were used against these attacks, at the time, were not
particularly impressive. {\em Logic and accuracy testing} (commonly
shortened to ``L\&A''), conducted prior to the start of the elction, would
run a small and pre-determined set of tests votes through the machine,
verifying that the proper tally appeared at the end. Of course, if the
machines watched their internal clocks, they could behave correctly while
under test and then be malicious only on Election Day. Similarly, the
number of votes used in L\&A is typically much smaller than will appear in
a real election, providing additional opportunities for a voting machine to
distinguish between test conditions and a real election, and thus behave
properly during L\&A.

A more aggressive mitigation, only used in a handful of jurisdictions, was to
conduct a {\em parallel test}. With a this, some fraction of the voting
machine population is randomly selected and then, rather than being
deployed to the field, is instead set up in the elections warehouse where
an operator enters a full day's worth of votes according to a script. As
with L\&A, the post-election tally from these machines under test has a
known correct outcome and any deviation from this would indicate a serious
problem. Because no real votes are being cast, video can also be captured
to ensure that operator data-entry errors can be differentiated from
malicious vote flips.

A thought experiment on how to defeat this, which can possibly be
attributed to Avi Rubin, is to have a {\em secret knock}. This is an input
that no rational testing process would ever contain, but which malware
would look for. The canonical example is a write-in vote for Mickey Mouse,
although that might well happen in practice, so an attacker would need to
select something more obscure. The general idea, then, is that the malware
will act identically to the legitimate software until it sees the secret
knock, and only then start misbehaving. This would be effectively
undetectable without potentially destructive forensic testing, although it
requires co-conspirators to perform the secret knock during the real
election, creating significant risks for the conspiracy.

\subsection{Would all these attacks and defenses still work in a BMD?}

This is the crux of ADS's argument. Malicious software in a BMD
can certainly show one thing on the screen and print something else on the
paper. This is particularly troublesome with some vendors who print two
different encodings of the ballot: barcodes for machine-readable data and
printed text for humans. While voters can verify the printed text, no voter
will be able to detect errors in the barcodes.  Let's call this an {\bf\em
  inconsistent barcode attack}. Alternately, the machine might produce a
completely consistent paper ballot (i.e., the barcode and the
human-readable text are in total agreement), but the paper ballot differs
from what the voter entered on the touchscreen. Let's call that a {\bf\em
  switched intent attack}.

There's a very simple solution to the inconsistent barcode attack:
{\em get rid of the barcodes} and have the only record of the voter's
intent be human-readable text. Any computer-printed text that's
readable by a voter will also be readable by a computer scanner with
exceptionally high accuracy\footnote{To achieve the necessary
  accuracy, the OCR software likely needs to use state-of-the-art
  ``deep learning'' techniques. Until such an OCR technology
  can be proven out in practice, barcodes will be necessary as a
  bridging technology.}. While many current BMDs do not operate this
way, this can be addressed through regulatory mandates and software
updates from the vendors.

\subsection{Auditing and inconsistent ballot detection}
So long as paper ballots have both human-readable text and a barcode, we
need to consider how existing audit processes can be adapted to detect
these attacks. A number of audit processes, including recounts and risk
limiting audits (RLAs), provide these opportunities.

If even a single ballot is inconsistent, that's evidence of a serious
problem---either a major software bug or an inconsistent-barcode
security attack---that would require emergency procedures (discussed below
in Section~\ref{sec:emergency}). Because
of this, inconsistent barcode attacks are unlikely to be
mounted by attackers in the real world, because even a single inconsistent
ballot represents incontrovertible evidence that something has gone wrong. 
Attackers who wish to quietly manipulate an election outcome would not
want to leave this kind of evidence so easily available for discovery.

What about a switched intent attack then? This seems preferable to an
attacker, since it's not immediately obvious when it occurs. How can we
discover one? ADS base their argument against such discoveries on
three factors: that voters are unlikely to notice these attacks, that even
if a voter does discover such an attack there is no good process to respond
to such discoveries, and that there's no alternative process in place that
might reliably detect the attacks.

\subsection{Will voters notice a switched intent attack?}
\label{sec:switched-intent}
A number of studies were conducted at
Rice~\cite{everett07thesis,campbell09}, where they create a paperless
voting machine that deliberately lied on its summary screen. Their
goal was to detect how many research participants, drawn from the
local population, would notice that the machine changed their inputs
and would then go back and fix their ``mistake.'' Depending on exactly
how they set up the experiment, between 1/3 and 1/2 of the voters
noticed the introduced errors. 

In most of these experiments, the participants were given made-up
names on a printed sheet and asked to vote for them. It's entirely
possible that with real candidate names, in a real election,
particularly at the top of the ticket where name recognition will be
higher, voters might be more likely to notice discrepancies. This
suggests that, if there were systemic vote flipping malware, something
that tried to move thousands or tens-of-thousands of votes, that we
would have large numbers of regular voters who recognize the error
when it happens.

(A more recent manuscript by DeMillo et
al.~\cite{demillo-verification-2018} describes two studies. The
first considers timing data from observed live voters. The second
presents results from an exit survey of voters, presented with
a blank ballot and asked if it was equal to the actual ballot they
voted. Neither of these were controlled studies, so
their observations are unreliable for predicting verification rates.)

What might happen if a voter notices an error on a printed BMD? Most
voters will likely head back to the poll workers' table, perhaps
sheepishly admit to having made a mistake, and request a chance to
repeat the process with a fresh ballot. This process, commonly called
{\em ballot spoiling}, is a completely standard part of any election
process involving paper ballots. In Texas, for example, a voter is
entitled to three attempts.

We can expect there to be a certain {\em background rate} of spoiled
ballots, no matter the correctness of the ballot marking devices, so
it's only when the spoilage rate gets reliably above the background
rate that we'll have a useful signal. Clearly, poll workers need to
track every time a voter spoils a ballot and election administrators
need this data available as the election is ongoing, giving them
real-time situational awareness of problems as they manifest.

Given all this information, what should an election official do when
the spoiled ballot rate is higher than expected? Preferably, they
would have a variety of different responses available, from deploying
additional auditors to more serious emergency procedures.
(We discuss the exact likelihood of this detection in Section~\ref{sec:spoiled}.)

\section{Live auditing of BMDs}
\label{sec:live-audit}
Election officials need procedures for conducting audits on BMDs, in the
field, while the election is ongoing. Because BMDs retain no internal
memory of cast votes, the only hard requirement for conducting any sort of
live audit is that any ballots printed during the audit must be kept out of
the ballot box. Such a process will be naturally transparent to voters or
election observers, who would be free to witness the process.

Who should conduct the audits? Audits might be conducted by poll
workers, as part of their regular duties, or they might be conducted by
dedicated auditors, working for the election administration, driving from
one polling location to another during the election period. The essential
attributes of a good auditing process are that ``enough'' tests are
conducted to observe rare events, and that these tests are sufficiently
random that a malicious BMD has no way to reliably determine whether
it is operating with a real voter or with an auditor.

If a BMD is going to misbehave, the auditor will have a chance to catch
it. And {\em if any auditor, anywhere in the county, catches even one
  malicious machine in the act, the game is over}. Call the police; we've
got evidence of a serious crime. (See Section~\ref{sec:emergency} for a
discussion of emergency procedures.)

This idea of live auditing has been around since at least Benaloh's
challenge mechanism~\cite{benaloh06simple,benaloh07evt}, quickly
adopted by research voting systems like
Helios~\cite{adida08helios,helios-louvain09} and
VoteBox~\cite{sandler07auditorium,sandler08votebox}. Even without the
cryptography, the concept is the same. We wish to test a machine to
prove that it's generating correct output. The machine doesn't know 
that it's being tested. The machine must {\em commit} to its output,
and then we can verify the correctness of that output, or
alternatively arrive at concrete proof of the machine's misbehavior.

\setlength{\tabcolsep}{12pt}

\begin{table}
\begin{center}
\begin{tabular}{rrr}
{\bf Prob. Cheating ($p$)} & {\bf Audits ($n$)} & {\bf Prob. Detection} \\
\hline
1\% &	40 &	33.10\% \\
	&80 &	55.25\% \\
	&120 &	70.06\% \\
	&160 &	79.97\% \\
	&200 &	86.60\% \\
	&240 &	91.04\% \\
	&280 &	94.00\% \\
	&320 &	95.99\% \\
	&500 &	99.34\% \\
\hline
5\% & 10 & 40.13\% \\
 & 20 & 64.15\% \\
 & 30 & 78.54\% \\
 & 40 & 87.15\% \\
 & 50 & 92.31\% \\
 & 60 & 95.39\% \\
\hline
10\% & 10 & 65.13\% \\
 & 20 & 87.84\% \\
 & 30 & 95.76\% \\
 & 40 & 98.52\% \\
 & 50 & 99.48\% \\
\hline
15\% & 10 & 80.31\% \\
 & 20 & 96.12\% \\
 & 30 & 99.24\% \\
 & 40 & 99.85\% \\
 & 50 & 99.97\% 
\end{tabular}
\end{center}
\caption{\label{tab:detection}Probabilities of discovering
  a cheating voting machine as a function of the odds of 
  cheating occurring ($p$) and the number of audits ($n$) conducted.}
\end{table}

\subsection{Baseline audits}
An election director must conduct some amount of auditing, no matter what,
and in the event suspiciously high spoiled ballot rates are reported, the
election director might adaptively deploy more auditors.

What is the probability of catching at least one malicious machine in
the act? The math is straightforward. Let's say that a malicious BMD
does a switched intent attack with probability $p$. A randomly audited
machine would then be caught cheating, again with probability $p$.
Equivalently, the BMD gets away with its malice with probability
$1 - p$. The probability of the BMD getting away with it after $n$
audits is then $(1 - p)^n$. Equivalently, the probability of detecting
the malware is $1 - (1 - p)^n$. Table~\ref{tab:detection} shows some
real numbers for $p$ and $n$.  Gilbert has also suggested an auditing
process like this~\cite{gilbert2019}.

From this, we can see that if the attacker wishes to modify only 1\% of
the ballots, an election official wishing to detect such an attack will
need to conduct somewhere above 300 audits. To achieve a 99\% confidence
of detecting the attack, 468 audits would be necessary.

Of particular note, the probability of detecting a switched-intent attack
has no dependency on the number of votes cast in the election. This means
that the {\em proportional} cost of reaching a given detection probability
shrinks as the voting population grows. 

\section{Non-uniform malicious behavior}
\label{sec:non-uniform}
After a Twitter discussion of this idea, Stark wrote two drafts
of an essay in response~\cite{stark2019reliable-prev,stark2019reliable}.
This section responds to some of his arguments.

Stark's strongest claim is that the adversary can be far more selective
about which voters to attack. For example, the attacker might only tamper
with ballots for voters who operate very slowly, far more slowly than any
auditor. Similarly, an attacker might only tamper with ballots for voters that use
button boxes, large fonts, or other accessibility features. 

Stark argues that the complexity of the variations that specify a given
cast ballot, including the votes themselves, the time of day, the amount of 
time spent, and so forth, create a highly dimensional space that cannot
be efficiently audited. In his words:

\begin{quote}
Live tests need to probe every subset of voter preferences, BMD settings, and voter
interactions with the BMD that could alter any contest outcome, and they need to probe
every such subset enough to have a high probability of detecting any changes to selections
in that subset.
\end{quote}

\subsection{Thought experiment: The shoulder-surfing auditor}
\label{sec:shoulder}

Consider the following thought experiment: we assign auditors to voters
selected at random from the general voting population. Those voters will
then do their normal voting process, but an auditor will watch them
and will double-check the veracity of the printed ballot. This means
that we're selecting from the distribution of real voters rather than
from the distribution of all possible voter attributes.

This thought experiment is equivalent to Stark's ``oracle bound'' 
(Section 4.4, page 9). He offers as an example an election with 20 BMDs,
each of which prints 140 ballots during an election, for a total of 2800 ballots,
of which 14 ballots have been altered by malware in the BMDs. He concludes that
the shoulder-surfing auditor would need to observe $n=539$ voters to
achieve a 95\% chance of detecting the malware. 
Stark uses the following equation to find $n$:
\[
\frac{2800-14}{2800} \cdot \frac{2799-14}{2799} \cdot \cdot \cdot \frac{2800 - (n - 1) - 14}{2800 - (n - 1)} \leq 0.05
\]
This equation models a sequence of shoulder-surfing audits. The fraction on the left is
the probability that the malware survives the first audit, which is
to say, the probability that the first selected ballot was clean, which
is slightly less than 1.0.
As we move to the right, we're computing the probability that the
first audit was clean {\em and} the second audit was clean, which
we're thus multiplying together. We're
interested in how far to the right we need to get before the malware
wins with probably less than 5\%. 

Is Stark's math correct? How else might we model this? Using the
previous equation, with $p = \frac{14}{2800} = 0.5\%$ and $n = 539$,
we derive a detection probability of 93.29\%. Stark's version is more
precise in its counting, representing an error in the math of
Section~\ref{sec:live-audit} of 1.71\%.  As the number of ballots cast
in the election grows, the results of the two equations will converge.

Stark's argument hinges on his selection of a very small election,
with only 2800 ballots cast while also selecting a very small fraction
of malware activity, 0.5\%. This essay's live auditing strategy
demands a fixed number of audits regardless of the total number of
ballots cast, much like risk limiting audits select a number of
ballots as a function of the margin of victory, not of the number of
ballots cast.  Stark's math is correct, but his example is
cherry-picked to represent the very worst possible case for a live
audit on a BMD election.

BMDs seem to be of great interest to large counties and states that
will use them to collect millions of votes, with smaller counties
often selecting paper ballots because they don't need BMD-only
features like support for multiple languages, accessibility features,
or thousands of distinct ballot styles.  Stark's numbers do not
reflect the relative effort of live audits as they might be conducted
by the kinds of election jurisdictions that are favoring BMDs.

\subsection{Realizing the auditing scheme}
\label{sec:realizing}

Before Stark leaves behind his ``oracle bound'' model, he dismisses
it as being ``impossibly optimistic''. We next consider how realistic
this model might be in practice.

While we cannot assign auditors at random to specific voters, we can
create a probabilistic model of how real voters will behave. There are
two parts to a model like this: voting preferences (i.e., what the
voter ultimately selects in each contest), and machine-observable
behaviors of the voter (e.g., how slow or fast the voter enters their
preferences).

With historical ballots in hand, we can easily construct a
probabilistic model that reflects how often voters will vote
straight-ticket for one party or split their ticket; we can
differentiate this by precinct, since we can examine prior cast
ballots on a precinct-by-precinct basis.  While we cannot connect
prior cast ballots with the exact time they were cast---the cast
ballots should have been randomized---we can still look at data from
the event logs to determine long voters took to create and cast their
ballots. For features that we cannot observe from event logs, such as
how many times a voter back-tracked and changed a preference, we can
estimate these features from usability experiments.

As such, the challenge for the auditor is to create a ``random voter''
model that reflects all the voters across each voting precinct,
including their preferences and behaviors. This then produces an
auditing script, which will express the votes to be selected and the
manner in which to do it. The script must also specify the time and
location of each voting machine to be audited.

Political consultants regularly produce detailed models of voter behavior,
typically used by ``get out the vote'' campaigns and other efforts to
influence voter behavior. Similar techniques could be used to create
the audit scripts. Making this model realistic, for example, recognizing
that slow input behaviors may correlate with age, and age may correlate
with party preferences, is a significant part of the challenge, but
this challenge is still a tractable engineering problem.

\subsection{Down-ballot tampering}
\label{sec:downballot}

Stark also suggests that an attacker might target a down-ballot race,
where the number of audits that include that race might be quite small.
For example, in the Harris County (Houston, Texas) general election in
November 2018, roughly 1.2 million ballots were cast for statewide contests\footnote{\url{https://harrisvotes.com/HISTORY/20181106/cumulative/cumulative.pdf}}. For contrast, 
voters in the City of Baytown, on the east side of Harris County, cast roughly 9,000 votes in their
propositions on the same ballot. 

The actual election results from Baytown weren't close at all. The closest contest 
was decided by a nearly 2:1 margin. A malicious attacker would have needed to
tamper with roughly 1,500 ballots to change its outcome. To be really sure, since
the attacker could never have had such a precise prediction in advance, the
attacker would have probably chosen at least 2,000 ballots to tamper. The odds of a county-wide
``shoulder-surfing'' auditor observing one of those tampered 2,000 voters, if 300 audits were
performed, would be roughly 40\%. In a hypothetical variant of Baytown with a tighter
election and only 400 votes tampered, the odds of observing a tampered vote would be only 
10\%.

These numbers suggest that an election official will need other techniques,
besides live audits, to detect focused down-ballot tampering. We discuss
other processes next.

\subsection{Spoiled ballot rates as a signal of problems}
\label{sec:spoiled}

Stark analyzes the question of how many spoiled ballots would be
necessary to represent a statistically significant signal over a base
rate of ``expected'' spoiled ballots. He notes that the answer is a
function of the number of BMDs used; it's actually a function of the
number of ballots cast, but the math is the same.  With an assumed 1\%
spoilage rate, and 200,000 cast ballots, we might normally expect 2000
spoiled ballots\footnote{Until we see large deployments of BMD voting
  system, we won't know the actual rate of spoiled ballots on these
  systems. This data is certainly important to collect and study.},
with 74 additional spoiled ballots being a signal that the spoilage
rate was outside of the 95\% confidence interval.

Stark uses an assumption that only 10\%
of voters would notice BMD-introduced errors, which contradicts other
usability studies that have found voters detect errors more frequently. Nonetheless,
with Stark's 10\% assumption, he concludes that the malware can get away
with roughly 730 attacks, corresponding to changing the margin of victory
by 0.73\%; this includes the computation that one tampered ballot causes
one candidate to lose a vote and the other to gain a vote.
The table below generalizes Stark's math to three different
election sizes and three different likelihoods that voters
will detect tampered ballots and spoil them. The reported
Margin\ $\Delta\%$ indicates the maximum change to a margin
of victory that malware might hope to accomplish without
being detected.


\begin{table}
\begin{center}
\begin{tabular}{rrr}
{\bf Detection\%} & {\bf Num ballots} & {\bf Margin $\Delta\%$} \\
\hline
10\% & 9,000 & 3.556\% \\
     & 200,000 & 0.740\% \\
     & 1,200,000 & 0.300\% \\
\hline
30\% & 9,000 & 1.185\% \\
     & 200,000 & 0.247\% \\
     & 1,200,000 & 0.100\% \\
\hline
50\% & 9,000 & 0.711\% \\
     & 200,000 & 0.148\% \\
     & 1,200,000 & 0.060\% \\
\end{tabular}
\end{center}
\caption{\label{tab:margin}Possible malware changes to the margin of
  victory as a function of the detection rate and number of ballots
  cast. Code to generate this table is presented in Appendix~\ref{sec:computing}.}
\end{table}

Table~\ref{tab:margin} shows the impact of increasing the size of the
electorate as well as increasing the likelihood that voters will
notice and spoil tampered ballots. For a county-wide contest in Harris
County, any malware attack that moved the margin of victory more than
0.3\%, even with Stark's most pessimistic assumption about voters
noticing erroneous ballots, would pass the 95\% confidence interval on
the expected ballot spoilage rate. With a more optimistic assumption
on voters spoiling these ballots, the malware could not hope to move
the margin of victory by more than six one-hundredths of a percent
without exceeding the 95\% confidence interval.

Consequently, for elections in large jurisdictions, real-time tracking
of ballot spoilage events should provide an effective mechanism for
election officials to detect switched-intent attacks. For much smaller
elections, however, such as Baytown's local election, a
switched-intent attack would seem to have more room to operate.

\section{Election procedures and emergencies}
\label{sec:tactical}
As discussed in Section~\ref{sec:switched-intent} and quantified in
Section~\ref{sec:spoiled}, election officials gain power from having
situation awareness of the rate of ballots being spoiled. With this,
election officials can attempt to intuit strategies being taken by
malware and adaptively create auditing strategies that might catch the
malware in the act. If, for example, there were an unusually high
spoiled ballot rate in Baytown, or if there were a hotly contested
race there with large political stakes \footnote{Baytown is adjacent
  to a number of petrochemical refineries, including one of the
  largest in the U.S., owned by Exxon-Mobil, so it's easy to imagine
  complicated politics between the citizens of the city and the
  industry around them.}, then the election official could choose to
deploy additional auditors to those areas, on top of the baseline of
countywide live audits and spoiled ballot tracking.

As such, the ``game'' becomes less like flipping coins and more
like playing poker. Statistics still play a role, but the players must
spend a significant amount of energy trying to intuit each others'
strategies, including bluffing and other forms of subterfuge.

Furthermore, this is a game where the attacker must move first,
committing to a malware attack that will have a specific impact on the
outcome (whether a switched-intent attack or something else). The
election official gets to make responsive moves up to the day of
election, after the voting machines are potentially beyond the reach
of the attacker.

In the wake of the 2016 election, the Department of Homeland Security
declared elections to be ``critical infrastructure''. One of the
consequences of this is the existence of the Elections Infrastructure
Information Sharing and Analysis Center
(EI-ISAC)\footnote{\url{https://www.cisecurity.org/ei-isac/}},
creating a structured program for the federal government to share
threat intelligence information with election officials, as well as
for election officials to aggregate and share such intelligence
amongst themselves. Certainly, the compressed timetable of elections
means that this sort of sharing needs to happen quickly, but if, for
example, the various Federal intelligence agencies reach a conclusion
that attacks are likely in specific states or localities, then EI-ISAC
provides the necessary infrastructure to disseminate this information,
allowing election officials to perhaps get ahead of the malware before
the election even begins.

\subsection{Responding to an emergency}
\label{sec:emergency}

Stark raises this point in his responses, and it's an important issue
to discuss. While the legal process for managing elections varies from
state to state, consider what happens when a natural disaster strikes
right before or during an election. Exactly this happened with
Hurricane Sandy, which struck the northeastern seaboard on October 29,
2012, causing notably large damage in New Jersey and New York, right
before a presidential election. In the wake of this storm, many
politicians recognized the need for emergency procedures (see,
e.g.,~\cite{sandy-nyt-2013}), and the National Association of
Secretaries of State began a push to get states to adopt laws and
procedures to take disasters into
account~\cite{nass-emergencies-2017}. Ultimately, a cyberattack on an
election can and should be treated much the same as a hurricane or
other natural disaster. If the scope and reach of a cyberattack is
large enough that the outcome of the election is in doubt, then
suitable disaster procedures would allow a governor to declare an
emergency and re-run an election, perhaps with a different voting
technology.

Note that modern elections are not actually finalized on the night of
the election, even though losing candidates will customarily concede
to the victors at the time. Instead, all elections have a canvass
period after the election. During this period, a wide variety of
activities occur, which includes processes like tabulating
vote-by-mail ballots and resolving provisionally cast ballots. (See,
e.g., California's canvass information
page\footnote{\url{https://www.sos.ca.gov/elections/official-canvass/}}.)
The canvass period is typically when a risk-limiting audit will be
conducted, and is also a suitable time for cyber-forensics to be
conducted on BMDs that were discovered during audits or simply flagged
by voters spoiling their ballots.

Still, once a vote has been tampered, you cannot determine the intent
of the voter. So what do you do? Procedurally, this should be no
different than a ballot box, or potentially a warehouse of every
ballot box, being lost or destroyed in a flood. It's an emergency, and
you need emergency procedures to resolve the problem. While it would
be politically sensitive to declare that a cyberattack damaged an
election and as such it had to be re-run, the likelihood of an
emergency response mitigates against the risks of cyberattacks. In
other words, if the attacker doesn't think they'll get away with it,
they're less likely to bother with the attack.

\section{Additional arguments}
\label{sec:additional}

A wide variety of other arguments have been made by Stark and others
against ballot marking devices. This section tries to respond to
these arguments.

\paragraph{The adversary will know the election official's auditing strategy,
giving an advantage.} As discussed in Section~\ref{sec:realizing}, the
election official must create an auditing strategy that mirrors the real-world
distribution of voter's preferences and behaviors. This will inevitably require
a sophisticated software tool that constructs an auditing script based on
real-world data taken from prior elections. If this tool and the data it
uses are open-source, then the only input that the adversary doesn't know is
the random seed that drives the production of the audit script. If the model
has weaknesses that don't capture real-world voter behavior, the adversary 
may further be able to take advantage of these weaknesses. On the flip side,
the human auditors executing the auditing script will inevitably add randomness
on their own, e.g., making errors with respect to the scripted vote
inputs and needing to correct those.

A more serious threat is that the computer generating the auditing script
is, itself, controlled by the same adversary. This can be partly mitigated by having
multiple, independent computers generating the audit script, with the
random seed produced by rolling physical dice. The resulting scripts should be
identical, otherwise we have evidence of malware, which again leads
us to emergency processes.

\paragraph{These auditing procedures would require significant
  additional staffing and training.} Let's again use the November 2018
election in Harris County, Texas, to get some realistic numbers.  In
this election, 1.2~million votes were cast across a fleet of roughly
10,000~voting machines.  In this election, roughly 29\% of ballots
were cast on Election Day, with 63\% of ballots cast during the early
voting period. The remaining 8\% were absentee postal mail ballots.
In the 2019 general election, there were 52 early voting centers
and on Election Day there were 750 local voting locations.\footnote{
I can't seem to find these numbers for the 2018 election, but they
should have been similar to 2019.}

Let's assume that an auditing script for 2020 would follow the same distribution of
votes that we saw in 2018, with the bulk of ballots cast in early voting
locations. If we had one audit team (perhaps two trained auditors)
per early voting location, we would then need 104 trained auditors. If
we want 500 live audits across the full election, then this will
average out to roughly six audits per early voting location, spread
across the two-week early voting period, i.e., less than one audit per
day. As such, the auditors would most likely be regular poll 
workers given extra training for the auditing function.

On Election Day, the same 52 teams of auditors could then be charged
with executing roughly three audits per team, representing a very
reasonable workload to achieve around 500 audits across the election.
Even in the case where the probability of a BMD cheating was a low
1\%, this audit would have a 99.34\% chance of detecting it (see
Table~\ref{tab:detection}).  It's also straightforward to see these
auditors given additional auditing tasks, adaptively, as described in
Section~\ref{sec:tactical}.

\paragraph{These auditing procedures depend on the margin of victory,
which is not known while the audits are under way. If the final margin
of victory is smaller than the margin used for the audit script, then
the audit process cannot reach a conclusion.} 

This paper's analysis shows that live audits plus live tracking of
spoiled ballots can drive down the available margin for malware to
tamper with an election result without being detected. This raises a
related question of whether an adversary might even attempt this sort
of malware-based attack when other attacks (e.g., tampering with the
voter registration database, or bombarding social media with
propaganda) might be more likely to succeed and have a larger impact.

Perhaps more importantly, once the margin of victory gets within 1\%
of the number of votes cast, a variety of other factors come in to
play, particularly around usability of the voting system (see,
e.g.,~\cite{everett08chi-dre-usability}). BMDs have the potential to
perform much better than hand-marked paper in these circumstances, as
we discuss next.

\section{Discussion and Conclusions}
\label{sec:conclusions}
BMDs give us the opportunity to build more sophisticated voting
systems with ``end to end'' security guarantees. While none of today's BMDs
have features like this, the research literature has a variety of designs
that give voters a ``receipt'' that allows them to prove that their vote
was correctly included in the final tally (i.e., ``counted as
cast''). There are also many clever techniques that can be used to audit
voting devices to catch them if they're cheating (i.e., verifying that
voters are ``cast as intended''). If we ever want to have e2e elections,
then we will likely require BMD-like devices which produce regular paper ballots
as well as computing the necessary cryptography. 

How realistic is it that e2e will make the jump from the research
literature to commercial production? To pick one example, Microsoft is
investing in an open-source toolkit called
ElectionGuard~\cite{burt-ms-2019}, and they've announced partnerships
with many of the vendors of election equipment. It's quite likely that
the next generation of BMDs, and perhaps even current-generation BMD
hardware with new software, will adopt these techniques.

The risks of malware in current-generation BMDs are non-trivial, but
they can be mitigated through human-centered ballot design, careful
auditing procedures, and suitable election emergency laws. They also
keep the door open to new cryptographic techniques, such as used in
ElectionGuard, that have the potentially to protect against a variety
of other election threats.

Unlike the paperless electronic voting systems that BMDs are being
purchased to replace, the paper ballots that come out of BMDs give us
the ability to consider the security procedures contemplated
here. BMDs are the best technology available today that
combines the security benefits of paper with the accessibility
benefits of computers.

\section*{Acknowledgements}

I would like to thank my colleagues for reviewing drafts of this paper
and offering a variety of constructive suggestions:
Claudia Acemyan, 
Ben Adida,
Mike Byrne,
Joseph Hall, 
Philip Kortum,
and
Whitney Quesenbery.

{\small
\bibliographystyle{habbrv}
\bibliography{citations,votebox}
}

\newpage
\appendix
\section{Computing detectable spoiled ballot rates}
\label{sec:computing}

This short Python program shows how to generate
Table~\ref{tab:margin}. In my attempts to reproduce Stark's
numbers, he suggested I use NumPy's {\tt poisson.ppf}---
the Poisson cumulative distribution function---which is more
accurate than the more commonly used normal approximation to the
Poisson distribution: $\mu + 1.96 \sqrt{\mu}$. (Wikipedia
explains where the number~1.96 comes from in this
equation\footnote{\url{https://en.wikipedia.org/wiki/1.96}}.)

\lstset{language=python,showstringspaces=false}
\begin{lstlisting}[frame=single]
from scipy.stats import poisson

print "\\begin{tabular}{rrr}"
print "Detection\% & Num ballots & Margin $\Delta\%$ \\\\"

expectedSpoilage = 0.01
for detectionFraction in [0.1, 0.3, 0.5]:
    print "\\hline"
    for electionSize in [9000, 200000, 1200000]:
        mu = electionSize * expectedSpoilage
        cd = poisson.ppf(0.95, mu)
        print "%d\\%% & %d & %.3f\\%% \\\\" % 
          (detectionFraction * 100, electionSize,
           200.0 * (cd - mu) / (electionSize * detectionFraction))

print "\\end{tabular}"
\end{lstlisting}

\end{document}